\documentclass[twocolumn,english,aps,reprint,prl,footinbib,amssymb,floatfix]{revtex4-1}
\usepackage[T1]{fontenc}
\usepackage[latin9]{inputenc}
\usepackage{amssymb}
\usepackage{graphicx}
\usepackage{esint}
\usepackage{subscript}
\usepackage{epstopdf}
\usepackage{eqnarray}

\makeatletter

\DeclareRobustCommand{\greektext}{%
  \fontencoding{LGR}\selectfont\def\encodingdefault{LGR}}
\DeclareRobustCommand{\textgreek}[1]{\leavevmode{\greektext #1}}
\DeclareFontEncoding{LGR}{}{}
\DeclareTextSymbol{\~}{LGR}{126}
\providecommand{\tabularnewline}{\\}

\makeatother

\usepackage{babel}
\begin{document}
\begin{abstract}
In this paper we use an ab-initio quantum transport approach to study
the electron current flowing through lithiated SnO anodes for potential
applications in Li-ion batteries. By investigating a set of lithiated
structures with varying lithium concentrations, it is revealed that
Li\textsubscript{x}SnO can be a good conductor, with values comparable
to bulk $\beta$-Sn and Li. A deeper insight into the current distribution
indicates that electrons preferably follow specific trajectories,
which offer superior conducting properties than others. These channels
have been identified and it is shown here how they can enhance or
deteriorate the current flow in lithiated anode materials.
\end{abstract}

\title{Electronic Properties of Lithiated SnO-based Anode Materials}

\author{Dominik Bauer{*}}

\affiliation{Integrated Systems Laboratory, Department of Electrical Engineering
and Information Technology, ETH Z\"urich, Gloriastrasse 35, 8092 Z\"urich,
Switzerland}

\author{Teut\"e Bunjaku}

\author{Andreas Pedersen}

\author{Mathieu Luisier}

\affiliation{Integrated Systems Laboratory, Department of Electrical Engineering
and Information Technology, ETH Z\"urich, Gloriastrasse 35, 8092 Z\"urich,
Switzerland}

\maketitle
\email[*{dobauer@iis.ee.ethz.ch}

\section{Introduction }

In our modern society the increasing usage of portable electronic
devices such as smartphones, tablets, or laptops, calls for the development
of enhanced rechargeable batteries. Similarly, the complete replacement
of conventional cars based on fossil fuel by electrical vehicles will
only become possible if the storage capacity of their battery significantly
improves \cite{liang2009graphene,wu2012designing}. Li-ion batteries
have emerged as one of the most suitable candidates in all these applications
due to their relatively large energy and power densities. Still, lots
of progresses remain to be done before achieving the performance limit
of the Li-ion battery technology. For example, graphite, which is
the most commonly used anode material with its specific capacity of
$372\,\mathrm{mAh/g}$, is outperformed by many other components in
terms of Li-storage \cite{li2015mos2,winter1999electrochemical,chao2011situ,meduri2009hybrid}.
This is the case of tin-based structures, whose storage capacity can
reach values up to $940\,\mathrm{mAh/g}$ \cite{idota1997tin,park2010li,tamura2003advanced}.
However, this feature comes with large volume expansions when inserting
lithium ions into SnO or SnO\textsubscript{2} \cite{liang2009graphene,besenhard1997will,brousse1998thin}.
There are different approaches to overcome the possible performance
degradation. One is the formation of a composite material by combining
the host material with carbon \cite{xu2013electrochemical,zhou2013binding,yu2009tin}.
Another consists in arranging the active anode material into nanostructures
\cite{bazin2009high,zhao2015significant,kamali2011tin}. For the latter,
it is not only essential to understand how nanosizing affects the
morphology of the anode, but also how efficiently the freed or captured
electrons can flow through the formed atomic systems. This work focuses
on the characterization of electrical transport through nanostructured
lithiated SnO.

Experimentally investigating the different lithiation levels of SnO
can be very time-consuming. A convenient alternative consists in using
ab-initio (from first-principles) simulations, where there is no need
to actually fabricate material samples. Such approaches, based on
fundamental physics, are now well-established due to the increase
in available computer power and the rapid improvements in computational
methods \cite{kohan2000first}. 

The structural variations of a hypothetical lithiated SnO set has
already been studied at the ab-initio level in Ref.~\cite{pedersen2015three}
and agrees well with experimental data in terms of volume expansion
\cite{ebner2013visualization}. Based on these simulations, it has been proposed that the lithiation
and delithiation of SnO is reversible between 2 and 6.5 lithium ions
per SnO unit, which organize themselves as Li layers. Starting from
the suggested irreversible Li\textsubscript{2}SnO crystal and continuously
increasing the Li concentration results in two phase transformations,
one where the Sn atoms get dissociated from the formed Li\textsubscript{2}O
matrix and another where Li ions are placed between Sn atoms and cause
a large volume expansion.

This paper therefore concentrates on another aspect, the electronic
properties of the anode material, which might be equally important
as the structural changes. The electron conductance is indeed a critical
factor that determines the efficiency of the battery charge and discharge
processes. Ideally, the electrical current, which is a measure of
the conductance, should reach magnitudes comparable to those of bulk
$\beta$-Sn and Li. Although the presented results are specific to the lithiated
SnO system, the proposed simulation methodology can be applied to
any electrode material and shed light on the interplay between morphological
and electronic transport effects.

\begin{figure}
\centering{}\centerline{\includegraphics[width=1\columnwidth]{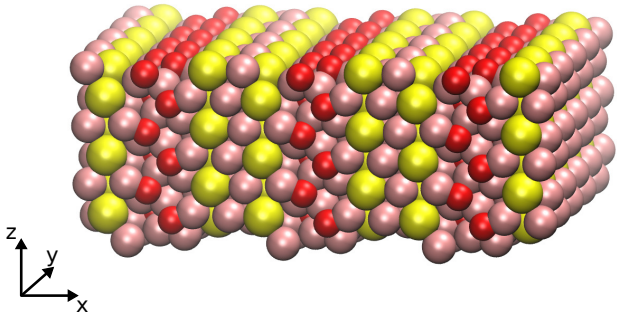}}\caption{Atomic structure of Li\textsubscript{4}SnO with four lithium ions
per SnO host unit. The pink, red, and yellow spheres correspond to
lithium, oxygen, and tin atoms, respectively.\label{fig:Side-view-of-Li4}}
\end{figure}

The paper is organized as follows. In Section~\ref{sec:Simulation-Approach},
the full simulation approach is described and explained. Section~\ref{sec:Results}
discusses the electron current flowing through the Li\textsubscript{x}SnO
structures ($2\leq\mathrm{x}\leq6.5$). The total current will be
analyzed first before insight into the preferred electron trajectories
is provided. Afterwards, a Bader charge analysis \cite{tang2009grid}
will be used to investigate the carrier distribution and derive its
influence on the current values. The last subsection focuses on the
fully lithiated Li\textsubscript{6.5}SnO configuration and its peculiar
characteristics.

\section{Simulation Approach\label{sec:Simulation-Approach}}

\begin{figure}
\centering{}\centerline{\includegraphics[width=1\columnwidth]{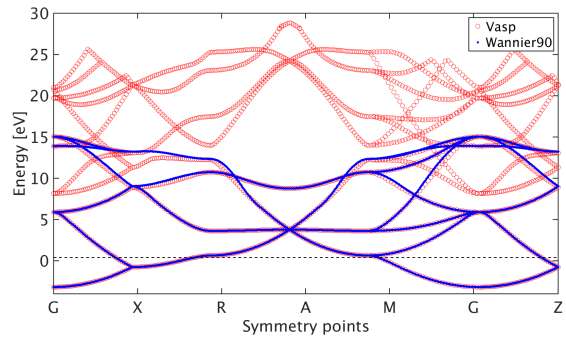}}\caption{Bandstructure of bulk Li at high symmetry points in the Brillouin
zone. The red curves refer to results obtained with VASP \cite{kresse1996efficient,kresse1996efficiency},
the blue curves with Wannier90 \cite{mostofi2008wannier90}. The dashed
black line indicates the Fermi energy. \label{fig:bs-w90-vasp}}
\end{figure}

To model the electronic properties of lithiated SnO anodes an approach
relying on density-functional theory (DFT) is used. It is based on
the plane-wave pseudopotential method, as implemented in the VASP
code \cite{kresse1996efficient,kresse1996efficiency}, within the
generalized gradient approximation \cite{perdew1996generalized},
as well as the projector-augmented wave formalism \cite{blochl1994projector}.
The number of k-points discretizing the Brillouin zone of each atomic
structure is chosen to be at least $3\times11\times11$. The plane-wave
kinetic energy cutoff is set to $600\,\mathrm{eV}$ with an electronic
convergence criteria of $10^{-6}$. The VASP calculations are performed
for the smallest possible rectangular unit cell that represents the
considered system, i.e. 4 atoms of tin and oxygen in the smallest
case and 16 in the largest.
The produced Bloch-Hamiltonian matrix corresponding to the chosen
periodic crystal is transformed into a set of maximally localized
Wannier functions (MLWFs) with the Wannier90 tool \cite{mostofi2008wannier90}.
This step is needed for transport calculations, as only a sparse Hamiltonian
matrix allows for the injection of electrons into a non-periodic simulation
domain. The accuracy of the MLWF transformation has been validated
by comparing the bandstructure obtained with VASP to that resulting
from Wannier90. As an illustration, Figure~\ref{fig:bs-w90-vasp}
shows the bandstructure of bulk Li. By keeping four Wannier functions
per Li atom (one with an s-like, three with a p-like symmetry), it
can be observed that around the Fermi level, both calculation methods
agree very well. Larger discrepancies are visible at higher energies,
but they can be ignored since those states do not contribute to the
transport properties.

The ballistic conductance of the Li\textsubscript{x}SnO anodes can
then be calculated in two ways. In the first one, the Landauer-B\"uttiker
formula \cite{buttiker1985generalized} is recalled to evaluate the
total current flowing through the anode material

\begin{equation}
I_{total}=\frac{2e}{\hbar}\sum_{k}\int\frac{dE}{2\pi}T(E,k)\left[f(E,E_{FL})-f(E,E_{FR})\right]\label{eq:landauer-new}
\end{equation}

where $e$ is the elementary charge, $\hbar$ the reduced Planck constant,
$T(E,k)$ the energy- and momentum-dependent transmission function
through the anode, $f(E,E_{FL/R})$ the Fermi-Dirac distribution function
of electrons in the added left (L) and right (R) contact regions,
and the factor 2 accounts for the spin degeneracy \cite{datta1997electronic}.
In Eq.~(\ref{eq:landauer-new}), a small bias difference $\triangle V\leq1\cdot10^{-3}\,\mathrm{V}$
is assumed to be applied between the left and right end of the Li\textsubscript{x}SnO
system. This value is appropriate for the short length of the studied
part of a potential full anode. To compute $T(E,k)$ at a given energy
$E$ and momentum $k$, the bandstructure of the primitive rectangular
unit cell can be employed. The latter can be rapidly determined in
the entire Brillouin zone with the generated MLWF Hamiltonian. Each
time a band with a positive velocity crosses the selected $(E,k)$
pair, $T(E,k)$ increases by one unit, i.e. if there are $n$ bands
fulfilling the required criterion, then $T(E,k)=n$. While this approach
might be computational very efficient, it does not offer any insight
into the trajectories that electrons follow through Li\textsubscript{x}SnO.
Only the current value is known.

To go one step further and identify the preferred current paths, a
ballistic quantum transport (QT) solver can be utilized. For that
purpose, the MLWF-Hamiltonian H\textsubscript{MLWF} of the primitive
rectangular unit cell is scaled up to enable the simulation of larger
structures. The resulting supercell matrix exhibits a block tri-diagonal
shape. Here, it is typically composed of four primitive cells along
the transport direction \cite{https://doi.org/10.3929/ethz-a-010659234}.

The QT calculations are performed within the wave function (WF) or
Non-equilibrium Green's function (NEGF) formalism after importing
the externally created MLWF Hamiltonian matrix \cite{luisier2006atomistic}.
In NEGF, the current between atoms $i$ and $j$ is given by

\begin{eqnarray} \label{eq:OMEN2}
I{}_{d,ij} & = & \frac{2e}{\hbar}\sum_{k,\sigma_{1},\sigma_{2}}\int\frac{dE}{2\pi}(H_{MLWF_{ij}}^{\sigma_{1}\sigma_{2}}(k)\cdot G_{ji}^{<\sigma_{2}\sigma_{1}}(E,k)\nonumber \\
 &  & -G_{ij}^{<\sigma_{1}\sigma_{2}}\cdot H_{MLWF_{ji}}^{\sigma_{2}\sigma_{1}}(k)).
\end{eqnarray}

In Eq.~(\ref{eq:OMEN2}), $H_{MLWF_{ij}}^{\sigma_{1}\sigma_{2}}(k)$
is the momentum-dependent Hamiltonian matrix element between atoms
$i$ and $j$ and between the Wannier functions $\sigma_{1}$ and
$\sigma_{2}$, while $G_{ij}^{<\sigma_{1}\sigma_{2}}(E,k)$ is the
lesser Green's function coupling atoms $i$ and $j$ as well as Wannier
functions $\sigma_{1}$ and $\sigma_{2}$ at energy $E$ and momentum
$k$. Note that because large supercells are constructed for transport,
the size of the Brillouin zone drastically decreases and no summation
over momentum is needed any more. The numerical problem reduces to
a \textgreek{G}-point calculation. The NEGF approach enables not only
to extract the local current between different orbitals, but also
includes the calculation of the total current via Eq.~(\ref{eq:landauer-new})
by indirectly delivering the transmission function $T(E,k)$. Both
of the above mentioned methods have been applied to our set of structures.
However, to consistently compare the total current values to the current
per bond analysis later on, the computation in this paper is restricted
to the quantum transport approach.

\section{Results\label{sec:Results}}

\subsection{Total current}

\begin{figure}
\centering{}\centerline{\includegraphics[width=1\columnwidth]{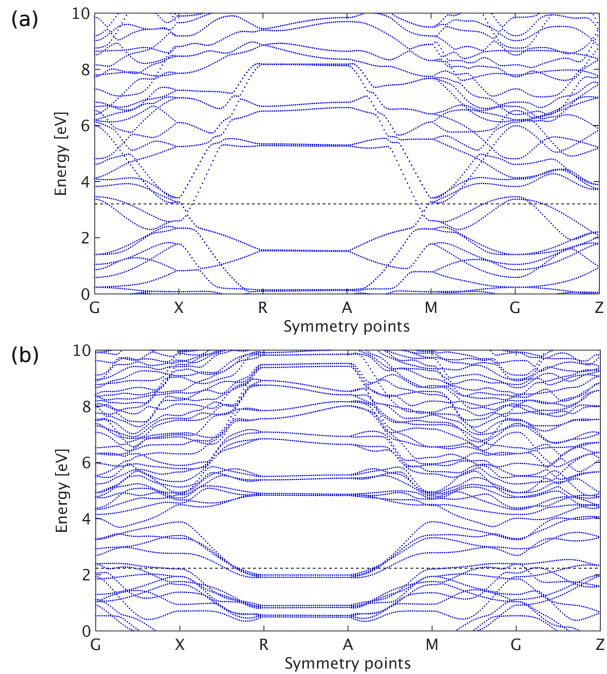}}\caption{Bandstructure of Li\textsubscript{4}SnO (a) and Li\textsubscript{6}SnO
(b) at high symmetry points in the Brillouin zone. The bands between
R and A correspond to transport along the x-axis. The dashed line
indicates the Fermi level of the respective structure. \label{fig:bs-w90-vasp_NEW}}
\end{figure}

\begin{figure}
\centering{}\centerline{\includegraphics[width=1\columnwidth]{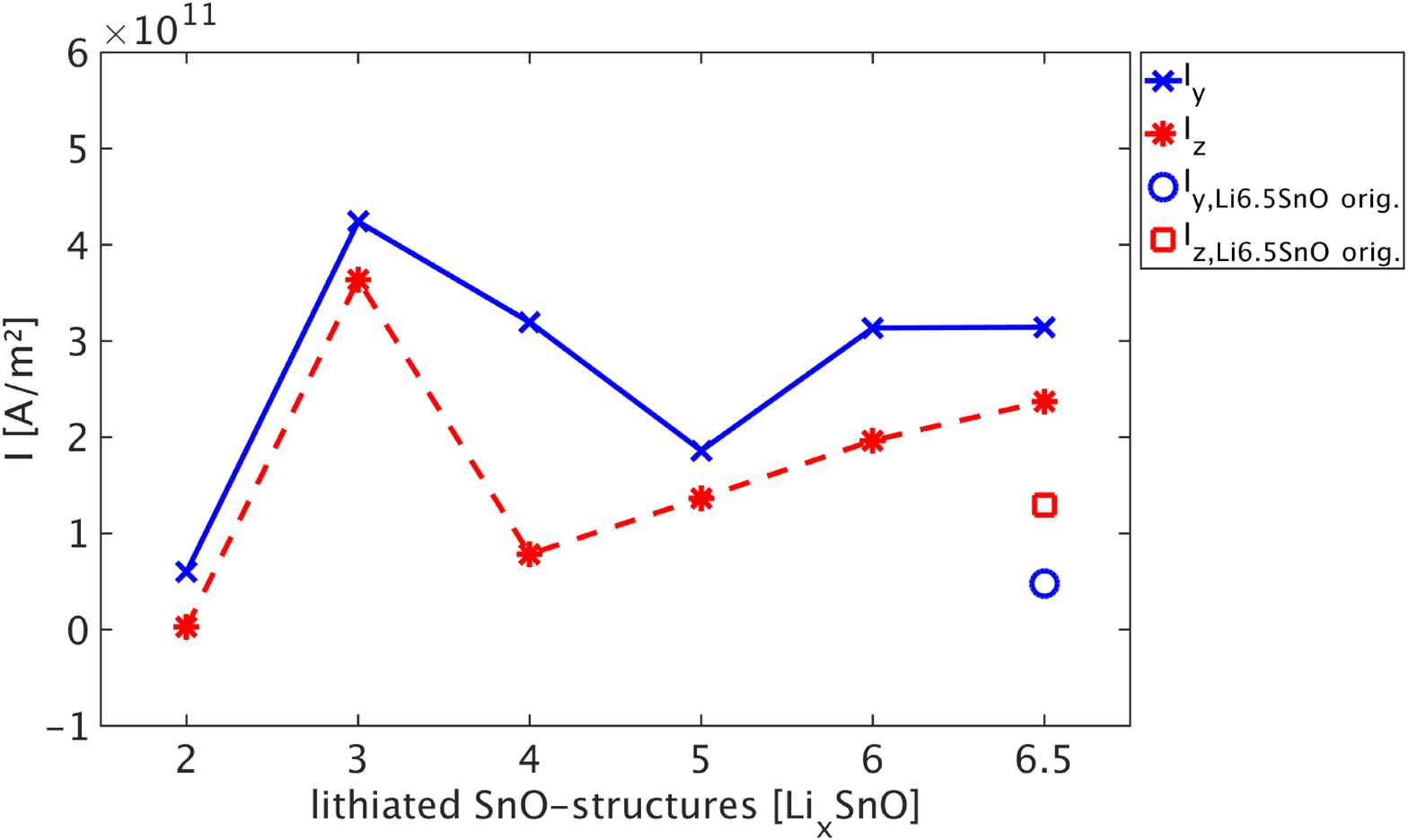}}\caption{Total current per area for the studied Li\textsubscript{x}SnO structure
measured from one contact to the other along the y- and z-directions
(see Fig.~\ref{fig:Side-view-of-Li4}), $I_{y}$ and $I_{z}$, respectively.
The x-direction is insulating because of the dense oxide layer separating
conductive segments. The horizontal axis shows the amount of lithium
ions per SnO host unit. For 6.5 layers, the two sets of markers (cross/asterisk
and circle/square) refer to the total current value obtained from
different configurations, as discussed in the last subsection of the
paper.\label{fig:The-total-current}}
\end{figure}

As a first step the total current flowing through the Li\textsubscript{x}SnO
structures is determined with Eq.~(\ref{eq:landauer-new}). Figure~\ref{fig:The-total-current}
reports the simulation results, both along the y- and z-axis, as defined
in Fig.~\ref{fig:Side-view-of-Li4}. The current along the x-axis
is omitted because the periodically repeated oxide layers along this
direction prevents any electron transport. Such structure can be clearly
seen in Fig.~\ref{fig:Side-view-of-Li4} for Li\textsubscript{4}SnO,
but is a common feature of all Li\textsubscript{x}SnO structures
with $\mathrm{2\leq x\leq6.5}$ Li layers. To further support this
statement, the bandstructures of all atomic systems have been investigated.
It has been found that all structures are metallic, except along the
k\textsubscript{x}-axis (R~-~A line in the Brillouin zone) where
the bands have a flat dispersion and are separated by a large band
gap, thus hindering electron transport along the x-axis and leaving
this direction insulating. Fig.~\ref{fig:bs-w90-vasp_NEW} illustrates
this behavior with two representative bandstructures, the one of Li\textsubscript{4}SnO
and Li\textsubscript{6}SnO. Orbital projected bands have also been
generated as an additional information, indicating that Sn d orbitals
are contributing the most to the bands around the Fermi level. When
$T(E,k)$ is evaluated with NEGF in Eq.~(\ref{eq:landauer-new}),
the computational burden becomes very high. It can be reduced by decreasing
the size of the supercell and the number of energy points, which introduces
a small error in the current of maximum $0.4\cdot10^{11}\mathrm{\, A/m^{2}}$.

The total current for bulk $\beta$-Sn and bulk Li has also been calculated
and compared to lithiated SnO. The values are $7.3\cdot10^{11}\,\mathrm{A/m^{2}}$
and $7.4\cdot10^{11}\mathrm{\, A/m^{2}}$, respectively. It can therefore
be concluded that Li\textsubscript{x}SnO as an anode material exhibits
good electronic properties, up to 60\% of those of bulk $\beta$-Sn and bulk
Li. Consequently, they are not expected to limit the charge and discharge
processes of a battery system with perfectly ordered electrodes. The
situation might change after few cycles, when different grained lithiation
levels start to be mixed together. The influence of disorder is however
out-of-the-scope of this paper.

It can be seen that the current in Fig.~\ref{fig:The-total-current}
does not follow a monotonic behavior with respect to the amount of
inserted lithium. At first, when going from 2 to 3 Li ions per SnO
unit, there is a large increase in current. According to Ref.~\cite{pedersen2015three}
this corresponds to the first phase transformation of the anode. When
inserting the first reversible Li layer into Li\textsubscript{2}SnO,
the oxygen region becomes fully filled up with lithium and the observed
Sn bilayer is transformed into a separate monolayer whose distance
to the O atoms is higher than in Li\textsubscript{2}SnO. This separated
Sn monolayer offers highly conductive paths, leading to an increase
in total current.

The second phase transformation occurs when a second reversible Li
layer is inserted, so transitioning from Li\textsubscript{3}SnO to
Li\textsubscript{4}SnO. The anode volume starts expanding in the
direction perpendicular to the oxygen plane, as lithium is now placed
in between the Sn atoms. This relocation introduces a larger distance
between neighboring Sn atoms and causes a decrease of the total current.
Along the y-direction, the effect is less pronounced because of the
presence of zig-zag Sn-channels that still provide a good conductance.
This will be discussed later. The larger drop in $I_{z}$ can be attributed
to the absence of such channels in the z-direction. When inserting
the fifth Li layer into SnO, $I_{y}$ keeps decreasing as more Li
atoms relocate between the Sn atoms, thus finally dissolving the Sn-channels.
In the other direction (z), the current magnitude reincreases because
lithium channels start to form. However, as the conductance per lithium
atom is lower than in the Sn-channels, the growth rate of the current
is not comparable to the Li\textsubscript{2}SnO - Li\textsubscript{3}SnO
transition. This finding might seem counterintuitive since the current
of both bulk $\beta$-Sn and bulk Li lie in the same range so that Sn- and
Li-based channels should offer similar conductance values. The reason
behind this surprising behavior will be addressed later on.

At the next lithiation levels, the current steadily increases, as
more Li atoms are available for transport. Note that the current values
depicted at 6.5 Li layers in Fig.~\ref{fig:The-total-current} correspond
to different Li\textsubscript{6.5}SnO configurations. Additional
knowledge about bond-resolved current trajectories are needed to explain
them. Until then, the focus is set on the rest of the structures.

\subsection{Current per bond }

\begin{figure}
\centering{}\centerline{\includegraphics[width=1\columnwidth]{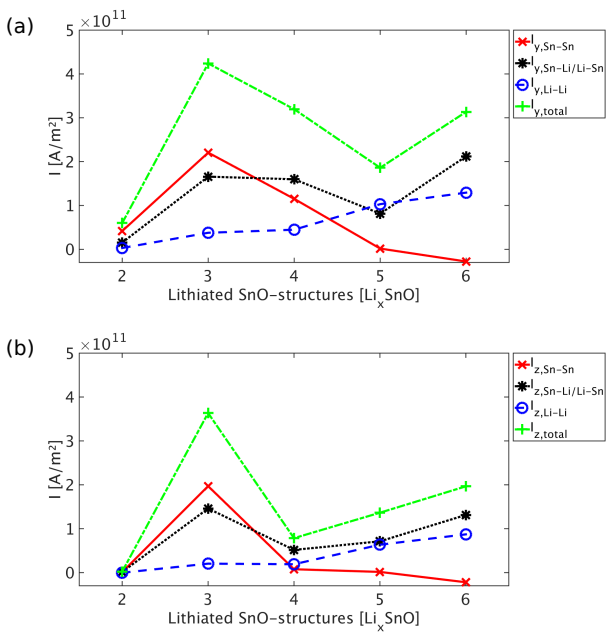}}\caption{Current components per area for each lithiated structure between Li\textsubscript{2}SnO
and Li\textsubscript{6}SnO along the y- (a) and z- (b) direction.
The same plotting conventions as in Fig.~\ref{fig:The-total-current}
are used. The red solid, black dotted, and blue dashed curves show
the current between Sn-Sn, Sn-Li/Li-Sn, and Li-Li atoms, respectively.
The green dash-dotted curve indicates the total current. \label{fig:The-current-components}}
\end{figure}

With the help of Eq.~(\ref{eq:OMEN2}) the current between two specific
atoms can be evaluated by determining all incoming and outgoing contributions
from their neighbors. After summing up those values the current components
from Sn-to-Sn, Li-to-Li, Sn-to-Li, and Li-to-Sn can be visualized
for the y- and z-direction, as reported in Fig.~\ref{fig:The-current-components}.
Both plots show similar results: following the first phase transformation
where Li\textsubscript{2}SnO evolves into Li\textsubscript{3}SnO,
the Sn atoms form separate channels, which is confirmed by the increased
Sn-to-Sn current. After this initial boost, the Sn-Sn current decreases
because the insertion of additional Li ions pushes the Sn atoms away
from each other, thus negatively affecting the Sn channels. This trend
continues as the lithium concentration increases. At Li\textsubscript{6}SnO,
the distance between neighboring Sn atoms is so large that the Sn-to-Sn
current contribution becomes negligible. Note that a negative current
is equivalent to reflections due to the absence of connections along
the transport direction. At the same time, the $\mathrm{I_{Li-Li}}$
component slowly increases due to the presence of more and more Li
ions in the anode and shorter distances between them. Thereby, the
conductance through Li atoms steadily substitutes that of Sn by forming
alternative channels. Hence, two opposite trends compete with each
other, the decrease of the Sn-to-Sn contributions and the increase
of the Li-to-Li ones. Since the latter dominates, the total current
increases between Li\textsubscript{5}SnO and Li\textsubscript{6}SnO.

Important information about the current flow resides in the $\mathrm{I_{Sn-Li}/I_{Li-Sn}}$
component. Figure~\ref{fig:The-current-components} reveals that
it has a high influence throughout the whole lithiation process. Hence,
even though direct connections between Sn atoms vanish, those particles
still significantly contribute to the overall current flow, but indirectly
via detours through Li atoms. 

\begin{figure}
\centering{}\centerline{\includegraphics[width=1\columnwidth]{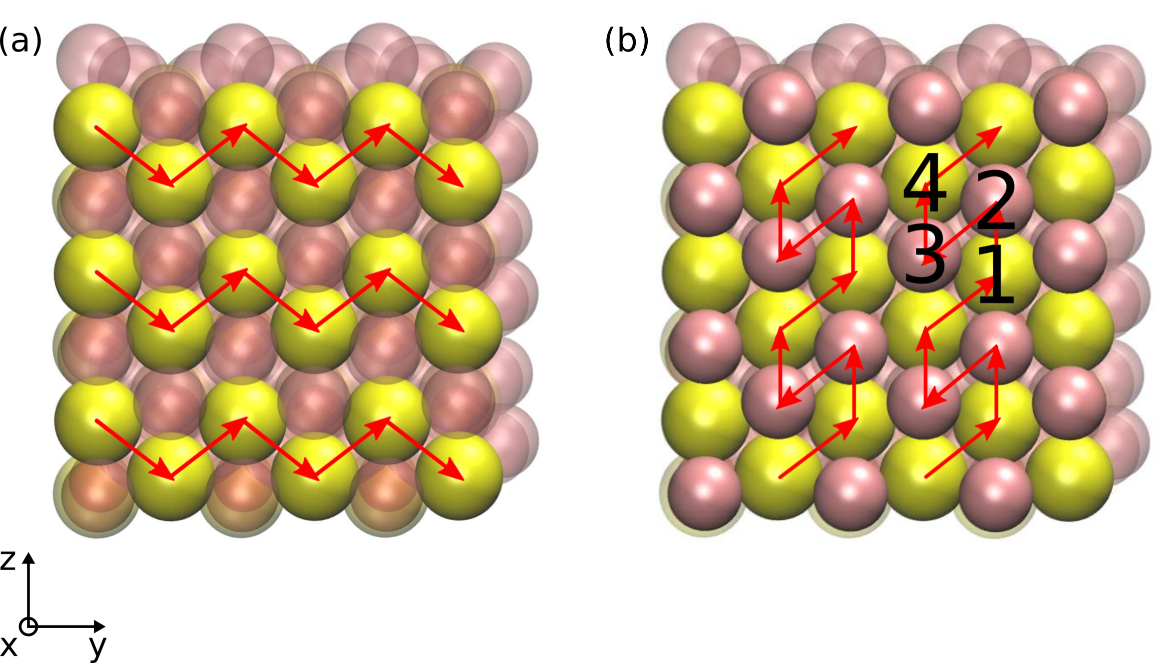}}\caption{Current map for Li\textsubscript{4}SnO along the y-(a) and z-(b)
direction. The dominant current flow is represented by the red arrows.
In (b), the current follows a non-direct path from atoms 1 to 4, always
hopping to the closest neighbor, i.e. the distance between atoms 1
and 2 is shorter than between 1 and 3. Similarly, atom 2 is closer
to atom 3 rather than 4.\label{fig:Current-path-in-Li4}}
\end{figure}

In the chosen MLWF approach one atom is connected to several neighbors,
but not all of them contribute equally to the current magnitude. By
only keeping the connections with the highest weight it is possible
to draw a trajectory map of the current. This is done by representing
the largest current components through a directional arrow that starts
from the atom position and points to the dominant direction. An example
is given in Fig.~\ref{fig:Current-path-in-Li4} for Li\textsubscript{4}SnO.
It should be noted that the current along the y-direction is significantly
higher than along the z-axis. This is due to the zig-zag alignment
of the tin atoms along y, which resembles the configuration of bulk
$\beta$-Sn and is therefore favorable to electron transport. Along z, the
current flow is perpendicular to these zig-zag channels so that the
distance between two Sn atoms situated in adjacent channels is too
large to allow for a direct connection. Thus, neighboring Li atoms
must be used as surrogate links between Sn channels. Since this process
is not as efficient as direct Sn-Sn coupling a decrease in total current
is observed, which explains the different behaviors along the y- and
z-direction mentioned in the previous section.

\subsection{Charge distribution }

Our simulations indicate that the electrical current flowing through
chains of Li atoms is considerably lower than through Sn channels,
whereas bulk Li and bulk $\beta$-Sn exhibit comparable current values. To
determine why Li and Sn atoms behave differently when they are mixed
together in Li\textsubscript{x}SnO structures, a Bader charge analysis
can be utilized \cite{tang2009grid}. It provides an accurate estimation
of the charge distributed on each atom based on the VASP outputs.
Table~\ref{tab:The-charge-distribution} summarizes the results for
each of the Li\textsubscript{x}SnO structure, as well as for bulk
Li and bulk $\beta$-Sn. 

\begin{table}
\begin{centering}
\centerline{{\scriptsize }%
\begin{tabular}{|c|c|c|c|}
\hline 
 & {\scriptsize Sn {[}e{]}} & {\scriptsize O {[}e{]}} & {\scriptsize Li {[}e{]}}\tabularnewline
\hline 
\hline 
{\scriptsize Li\textsubscript{2}SnO} & {\scriptsize $4.15$} & {\scriptsize $7.50$} & {\scriptsize $0.16$}\tabularnewline
\hline 
{\scriptsize Li\textsubscript{3}SnO} & {\scriptsize $4.80$} & {\scriptsize $7.70$} & {\scriptsize $(0.17$~-~$0.20)$}\tabularnewline
\hline 
{\scriptsize Li\textsubscript{4}SnO} & {\scriptsize $5.60$} & {\scriptsize $7.70$} & {\scriptsize $(0.16$~-~$0.19)$}\tabularnewline
\hline 
{\scriptsize Li\textsubscript{5}SnO} & {\scriptsize $(5.60$~-~$7.20)$} & {\scriptsize $7.70$} & {\scriptsize $(0.16$~-~$0.19)$}\tabularnewline
\hline 
{\scriptsize Li\textsubscript{6}SnO} & {\scriptsize $7.20$} & {\scriptsize $7.70$} & {\scriptsize $(0.17$~-~$0.20)$}\tabularnewline
\hline 
{\scriptsize Li\textsubscript{6.5}SnO \textsubscript{orig.}} & {\scriptsize $(7.50$~-~$7.60)$} & {\scriptsize $7.70$} & {\scriptsize $(0.17$~-~$0.20)$}\tabularnewline
\hline 
{\scriptsize Li\textsubscript{6.5}SnO \textsubscript{adj.}} & {\scriptsize $(7.20$~-~$7.80)$} & {\scriptsize $7.70$} & {\scriptsize $(0.17$~-~$0.23)$}\tabularnewline
\hline 
{\scriptsize bulk Li} & {\scriptsize $0$} & {\scriptsize $0$} & {\scriptsize $1$}\tabularnewline
\hline 
{\scriptsize bulk $\beta$-Sn} & {\scriptsize $4$} & {\scriptsize $0$} & {\scriptsize $0$}\tabularnewline
\hline 
\end{tabular}}
\par\end{centering}

\caption{Electron charge distribution in each of the considered Li\textsubscript{x}SnO
structures as well as in bulk Li and bulk $\beta$-Sn. Two Li\textsubscript{6.5}SnO
configurations, one labeled original (orig.), the other adjusted (adj.)
structure, are presented. The difference between them will be discussed
in Section~\ref{sub:LiSnO:-original-structure}.\label{tab:The-charge-distribution}}

\end{table}

When comparing the bulk to the Li\textsubscript{x}SnO structures,
it appears that in the latter case, the charge distribution strongly
influences the electronic transport properties. The highly reactive
Li ions, on average, give away more than 80\% of their charge to the
Sn and O atoms. In the Li\textsubscript{2}SnO case, the two Li layers
are tightly bound to the oxygen atoms, to which they both transfers
$0.75$~electron to form an irreversible Li\textsubscript{2}O layer.
The Li ions that are subsequently added to Li\textsubscript{2}SnO
donate roughly $0.8$~electron to their neighboring Sn atoms. As
a consequence of these charge transfer processes, the conductance
of the Li ions decreases.

\begin{figure}
\centering{}\centerline{\includegraphics[width=1\columnwidth]{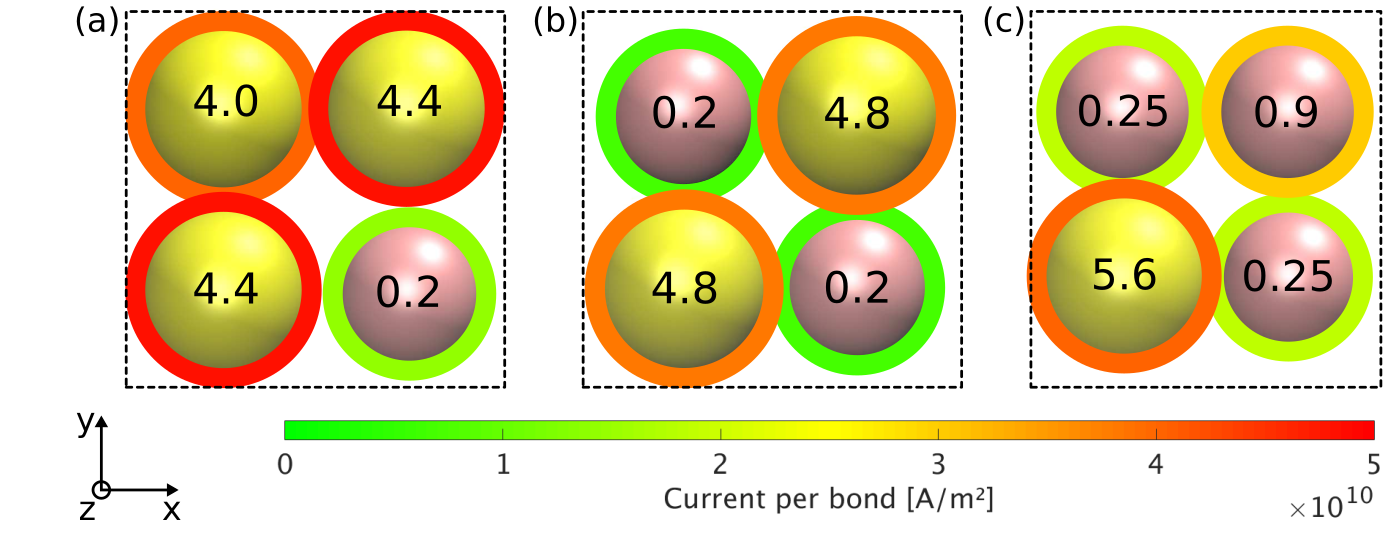}}\caption{$\beta$-Sn structures where one (a), two (b), and three (c) tin atoms have
been replaced by a Li atom. The numbers on the atoms show the calculated
electron charge. The colored circles around the atoms refer to the
computed current magnitude per bond along the z-direction. The dashed
rectangle represent the primitive unit cell. All of the displayed
atoms lie in the same z-plane. \label{fig:Li-into-sn}}
\end{figure}

The correlation between the charge and conductance variations is not
obvious at all and requires therefore a careful inspection. This is
done by investigating a simple composite SnLi system. A primitive
unit cell of bulk $\beta$-Sn with four atoms is considered. A Bader charge
analysis is performed and the current per bond is simulated after
replacing the Sn atoms one by one with Li atoms. The relaxed configurations
have only negligible bond connections between dissimilar atom types
along the z-direction. In this way, pure lithium and tin atomic channels
can be created and their current compared to each other. The results
are presented in Figure~\ref{fig:Li-into-sn}. In subplot~(a), the
inserted Li donates $0.8$~electron to the two adjacent Sn atoms,
whose charge becomes $4.4$~electrons each. The last Sn atom keeps
its original charge ($4$~electrons).

In this configuration, the current per bond that flows through the
pure Li channel along the z-axis is equal to $1.4\cdot10^{10}\,\mathrm{A/m^{2}}$
and is much smaller than for the Sn-Sn cases ($4.8\cdot10^{10}\,\mathrm{A/m^{2}}$
for the atoms with $4.4$~electrons, $4.0\cdot10^{10}\,\mathrm{A/m^{2}}$
for those with $4.0$~electrons). This is a first indication that
the charge transfer to neighboring Sn atoms negatively affects the
Li conductance, while it benefits to the Sn channels whose current
increases.

This trend is confirmed in Fig.~\ref{fig:Li-into-sn}(b) where again
the Li atoms give away $0.8$~electron to the Sn elements sitting
in their immediate vicinity and thus exhibit much lower currents ($0.7\cdot10^{10}\,\mathrm{A/m^{2}}$
for the Li-Li channels vs. $3.8\cdot10^{10}\,\mathrm{A/m^{2}}$ for
the Sn-Sn ones). Finally, the same observation can be made in Fig.~\ref{fig:Li-into-sn}(c).
The Li atoms that have been added all donate a fraction of their charge
to the Sn atom they surround, but not in the same proportion. Essentially,
the particle which lost only 10\% of its charge carries a higher current
($3.0\cdot10^{10}\,\mathrm{A/m^{2}}$) than the ones which gave away
$0.8$~electron ($I_{d}=1.8\cdot10^{10}\,\mathrm{A/m^{2}}$). Furthermore,
these current magnitudes remain much lower than in the Sn-Sn channels
($4.0\cdot10^{10}\,\mathrm{A/m^{2}}$). From these three numerical
experiments detailed in Fig.~\ref{fig:Li-into-sn} it can be definitively
concluded that when Li atoms sit close to Sn ones, a transfer of charge
takes place between both atom types. Thereby, the charge loss leads
to a current reduction in the Li-Li channels.

\subsection{Li\textsubscript{6.5}SnO: original structure vs. adjusted configuration\label{sub:LiSnO:-original-structure}}

\begin{figure}
\centering{}\centerline{\includegraphics[width=1\columnwidth]{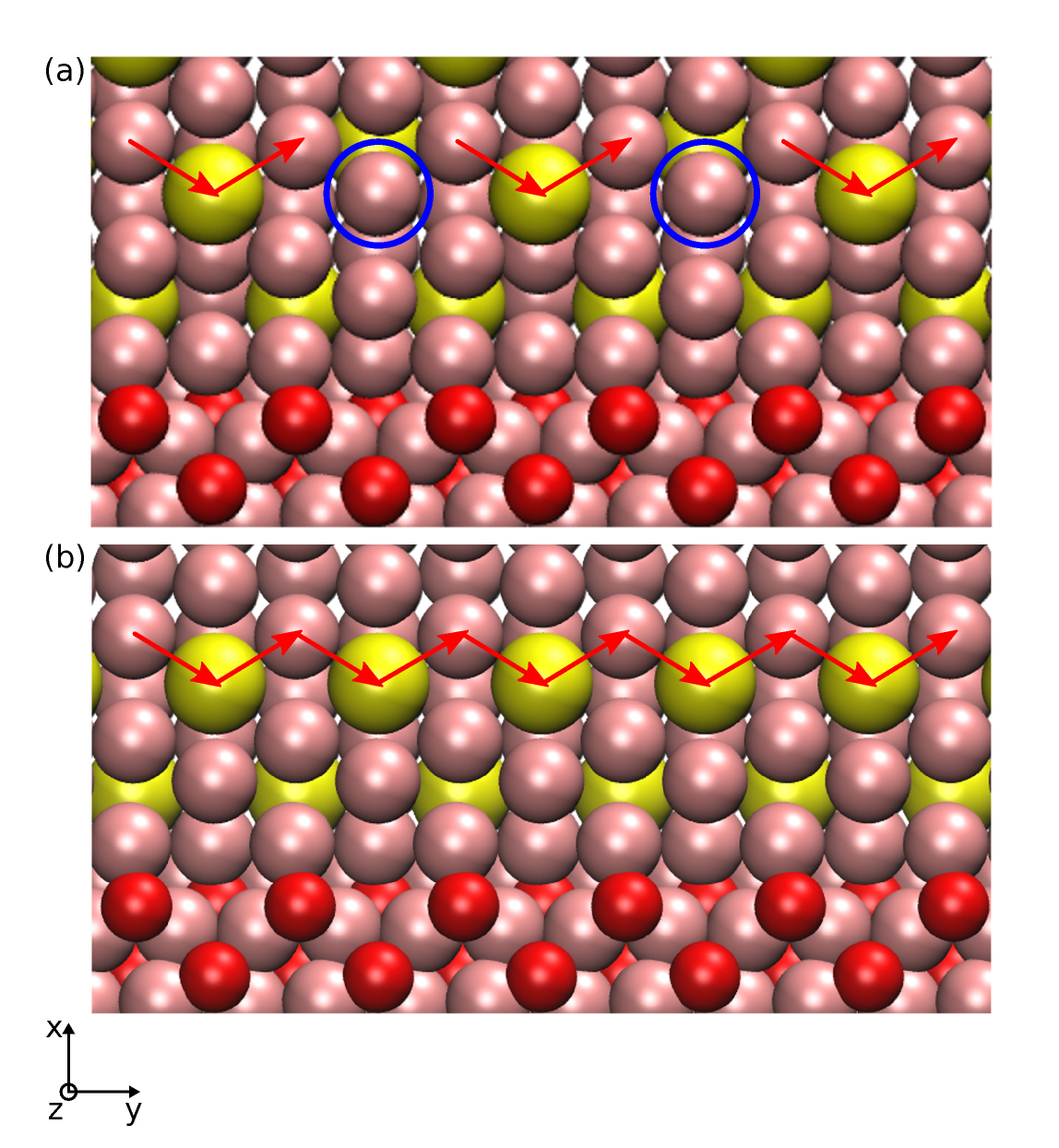}}\caption{Atomic configuration and dominant current path for the original (a)
and the adjusted (b) configuration of Li\textsubscript{6.5}SnO. In
(a), the blue circles refer to lithium spots, which are replaced by
Sn atoms in the adjusted configuration to form alternating Sn-Li-Sn
channels. \label{fig:Li6.5-current-path}}
\end{figure}

When simulating the original Li\textsubscript{6.5}SnO structure found
in Ref.~\cite{pedersen2015three}, it has been discovered that the
total current suddenly drops, as compared to Li\textsubscript{6}SnO.
This unexpected behavior is shown in Fig.~\ref{fig:The-total-current}.
The original Li\textsubscript{6.5}SnO configuration relaxes in such
a way that no alternating Sn-Li-Sn channels are formed, as illustrated
in Fig.~\ref{fig:Li6.5-current-path}(a). However, by studying the
lower lithiated arrangements and applying the lessons learnt in these
cases, an ``adjusted'' structure with more favorable current paths
can be created and relaxed according to the prescriptions from Ref.~\cite{pedersen2015three}.
Its total energy is 0.18\% higher than in the original structure with
a similar charge distribution, as can be seen in Table~\ref{tab:The-charge-distribution}.
Altogether, this makes it a possible alternative configuration.

Through transport simulations the dominant current trajectories can
be again identified. In Figure~\ref{fig:Li6.5-current-path}, the
focus is set on the differences between the original and the adjusted
structure. In subplot~(a), it can be observed that the missing Sn
atoms at the circle locations do not allow for the completion of the
Sn-Li-Sn channels that carry most of the current in the Li\textsubscript{5}SnO
or Li\textsubscript{6}SnO structures. With the atomic rearrangement
in Fig.~\ref{fig:Li6.5-current-path}(b), these channels reappear,
which increases the total current value. Hence, the trend that started
with the lesser lithiated SnO materials continues, as depicted with
the cross/asterisk markers in Fig.~\ref{fig:The-total-current}.
The unfavorable current flow in the original structure could come
from the fact that the lithium load has already reached its theoretically
predicted limit of Li\textsubscript{4.4}Sn (+ Li\textsubscript{2}O)
\cite{courtney1997electrochemical}. It has been verified that forcing
a similar atomic arrangement as in Fig.~\ref{fig:Li6.5-current-path}(a)
for lower lithiated structures, e.g. Li\textsubscript{6}SnO or Li\textsubscript{5}SnO,
does not reduce their total energy.

\section{Conclusion}

We have investigated the electronic properties of perfectly ordered
lithiated SnO at different lithium concentrations with the help of
density-functional theory and quantum transport simulations. This
work has mainly shed light on the influence of the atomic arrangements
on the total electrical current by evaluating the flow along each
bond connecting two atoms as well as by determining the charge distribution
in composite systems. The interplay between pure Sn, pure Li, and
mixed Sn-Li channels has been analyzed to explain the current behavior
as a function of the Li level. All in all, it has been found that
favorable current paths exist in Li\textsubscript{x}SnO with conductance
values close to what can be obtained in bulk $\beta$-Sn and bulk Li, except
in the direction orthogonal to the Li\textsubscript{2}O layers that
are created in the first irreversible lithiation step. This confirms
that SnO is a promising anode material, provided that the significant
volume changes it undergoes can be well managed. More generally, the
proposed approach to study electronic transport in battery electrodes
can now be used to explore other atomic systems.

\section*{Acknowledgments}

This research was supported by the European Research Council under
Grant Agreement No 335684-E-MOBILE.

\bibliographystyle{apsrev4-1}
\bibliography{re_revised_manuscript}

\end{document}